\documentclass{elsart}
\usepackage[english]{babel}
\usepackage{amsmath}
\usepackage{amsfonts}
\usepackage{amssymb}
\usepackage{graphicx}%
\newtheorem{theorem}{Theorem}

\newtheorem{proposition}[theorem]{Proposition}
\newenvironment{proof}[1][Proof]{\noindent\textbf{#1.} }{\ \rule{0.5em}{0.5em}}

\usepackage{amssymb}

\begin{document}

\begin{frontmatter}



\title{Extreme shock models: \\an alternative perspective}


\author{Pasquale Cirillo}
\ead{pasquale.cirillo@stat.unibe.ch}
\and
\author{J\"urg H\"usler}
\ead{juerg.huesler@stat.unibe.ch}

\address{Institute of Mathematical Statistics and Actuarial Sciences, University of Bern}

\begin{abstract}
Extreme shock models have been introduced in Gut and H\"usler (1999) to study systems that at random times are subject to shock of random magnitude. These systems break down when some shock overcomes a given resistance level.\\
In this paper we propose an alternative approach to extreme shock models using reinforced urn processes. As a consequence of this we are able to look at the same problem under a Bayesian nonparametric perspective, providing the predictive distribution of systems' defaults.\\
\end{abstract}


\end{frontmatter}

\section{Introduction}
Shock models are a particular class of models in which a system is randomly
subject to different shocks of random magnitude that can make it fail.\\
In the literature (see \cite{Huesler} for a good survey) there are essentially
two distinct types of shock models: cumulative shock models, in which the
failure of the system is due to a cumulative effect; and extreme shock models,
whose default is caused by one single extreme shock. In this paper we focus our attention only on the second type, referring to \cite{Gut2} for cumulative shock models.\\
Consider a family $\left\{  Z_{i},U_{i}\right\}  _{i\geq0}$ of nonnegative
i.i.d.\ two-dimensional random vectors, such that $Z_{i}$ represents the
intensity of the $i$-th shock and $U_{i}$ is the time between the $(i-1)$-th
and the $i$-th shock. Set $Z_0=U_0=T_0=0$ and, for $n\geq 1$, $T_n=\sum_{i=1}^n U_i$ is the amount of time that has elapsed after $n$ shocks.\\
In extreme shock models, one is interested in the behavior of the stopping time $\tau\left(  t\right)  =\min\left\{n:Z_{n}>t\right\} $, i.e. the minimum time at which a large extreme shock occurs causing the default of the system at $T_{\tau_t}$. We refer to  \cite{Gut4} for further details and results.\\
More recently, in \cite{Huesler}, extreme shock models have been generalized by assuming that  large but not fatal shocks may effect system's tolerance to
subsequent shocks. To be more exact, for a fixed $t$, a shock $Z_{i}$ can
damage the system if it is larger than a certain boundary value $\beta_{t}<t$.
As long as $Z_{i}<t$ the system does not fail. The crucial hypothesis is the
following: if a first nonfatal shock comes with values in $\left[  \beta
_{t},t\right]  $ the maximum load limit of the system is no more $t$, but
decreases to $\alpha_{t}(1)\in\left[  \beta_{t},t\right]  $, since the system
has been damaged. At this point, if another large but not too strong shock
occurs in $\left[  \beta_{t},\alpha_{t}(1)\right]  $, the new crucial
threshold is lowered again to $\alpha_{t}(2)\in\left[  \beta_{t},\alpha
_{t}(1)\right]  $ and so on until the system fails. We could call all this
\textquotedblleft risky threshold mechanism\textquotedblright. The relevant stopping time is now $\tau(t)=\min\left\{  n:Z_{n}\geq\alpha_{t}%
(L_{t}(n-1))\right\}  $ with $L_{t}(n)=\sum_{i=1}^{n}1_{\left\{  Z_{i}%
\geq\beta_{t}\right\}  }$ and $L_{t}(0)=0$. All the results can be found in \cite{Huesler}.\\
In this work we want to present an alternative perspective to extreme shock model by using a particular class of reinforced urn processes, as introduced by \cite{Muliere}. The aim is to develop a Bayesian nonparametric approach to shock models, in the wake of some recent works like \cite{Cihu}. As we will see the choice of using urn schemes for modeling is strictly related to their ability of reproducing the Bayesian paradigm of information update in a rather intuitive way.\\
In the literature there are very few papers that combine shock models and urn processes. In \cite{Marshall} a Polya urn is used to model a system subject to non-i.i.d. shocks, while in \cite{Cihu} a first urn-based approach to generalized extreme shock models is discussed.\\
Urn processes form a very large family
of probabilistic models in which the probability of certain events is
represented in terms of sampling, replacing and adding balls in one or more
urns or boxes. A good introduction to urn models and their properties is represented by \cite{Johnson} and \cite{Mahmoud}.\\
In reality it is not difficult to derive simple shock models using urns, as we only need basic tools as Bernoulli and Poisson trials. Think for example of a simple urn containing $b$ black balls and $w$ white balls, which is sampled with replacement. If black balls are associated to the event ``extreme fatal shock" and white balls represent the case in which there is no dangerous shock, then the waiting time related to the extraction of the first black ball from the urn is somehow related to the $\tau(t)$ in extreme shock models.\\
Naturally, this is just a trivial example. A more interesting urn scheme has been proposed in \cite{Cihu} to model generalized extreme shock models, where a triangular reinforcement matrix is used to mymic the risky threshold mechanism. This model is called UbGesm (urn-based generalized extreme shock model) and, in the next sections, we will show how to obtain it as a special case of our new approach.\\ 
Very briefly, having in mind the mechanism of generalized extreme shock models, the idea is to create three different risk areas for the system subject to shocks - no risk or safe, risky and
default, to link every area to a particular color and to work with the probability for the process to enter each area.\\
If every time the process enters the risky area the probability of failing increases, and this can be obtained with a triangular reinforcement matrix, we can consider such a modeling a sort of intuitive approach
to generalized extreme shock models, getting around the definition of the moving threshold. In some
sense, reinforcing the probability for the system to fail is like making the
risky threshold move down and vice-versa. \\
Consider an urn containing balls of three different colors: $x$, $y$,
and $w$. The $x$-balls are related to the safe state, $y$-balls to the risky state
and $w$-balls embody the default state. The process evolves as follows: at time $n$ a ball is sampled from the urn, the probability of sampling
a particular ball depending on the urn composition after time $n-1$; then, according to the color of the sampled ball, the process enters (or remains in)
one of the three states of risk (e.g. if the sampled ball is of type
$x$, the process is in a safe state, while it fails if the chosen ball is $w$); finally the urn is reinforced according to its balanced reinforcement matrix, as shown in (\ref{urna}). It means that if the  sampled ball is of type $x$, then the ball is returned and $s>0$ extra $x$-balls are added to the urn, if the sampled ball is of type $y$, then it is replaced together with $r>0$ $y$-balls and $p>0$ $w$-balls, and if the sampled ball is of type $w$, then $s$ extra $w$-balls are added. 
\begin{equation}
RM=%
\begin{array}
[c]{c}%
x\\
y\\
w
\end{array}
\left[
\begin{array}
[c]{ccc}%
1+s & 0 & 0\\
0 & 1+r & p\\
0 & 0 & 1+s
\end{array}
\right]  ,\text{ where }r=s-p\label{urna}%
\end{equation}

The distributions of the different colors and the main properties of the urn process can be described analytically through the analysis of its generating function, as discussed in \cite{Cihu}, where all the connections with standard shock models are also analyzed.

\section{Extreme shock models via reinforced urn processes}
The basic ingredient of our construction is the Polya urn, introduced by \cite{Polya}. \\
The behavior of this urn model is very simple yet ingenious. In its simplest version, we have an urn containing balls of two different colors. Every time we sample the urn we look at the color of the chosen ball and then put it back into the urn together with another ball of the same color. In this way, the more a given color has been sampled in the past, the more likely it will be sampled in the future. It is easy to understand how this naive replacement rule is able to reproduce a lot of self-reinforcing and contagious phenomena.\\
More formally, at time $t=0$, consider an urn $U$ with initial composition $C=(w_0, b_0)$, where $w_0\geq 0$ and $b_0 \geq 0$ are the numbers of white and black balls. At every time step $t \geq 1$, urn $U$ is sampled and the extracted ball is replaced into it together with $s>0$ balls of the same color. Let $X_t$ be a random variable equal to 1 if the sampled ball at time $t\geq 1$ is black and 0 whether the ball is white. It is clear that $X_1\sim Bernoulli(Z_0)$, where $Z_0=b_0/(w_0+b_0)$ is the proportion of black balls at time $t=0$, and, for $t\geq 1$, $X_{t+1}\sim Bernoulli(Z_t)$, where $Z_t=b_t/(w_t+b_t)$ is now a random variable. It is not difficult to demonstrate that the sequence $\{X_t\}_{t\geq 1}$ is exchangeable in the sense of de Finetti \cite{deFinetti2}.\\
Regarding the evolution of balls in the urn, we simply have
\begin{equation}
(w_{t+1},b_{t+1})=
\begin{cases}
(w_{t},b_{t}+s) &\text{if }X_{t}=1\\
(w_{t}+s,b_{t}) &\text{if }X_{t}=0\\
\end{cases}.
\end{equation}
Thanks to its reinforcement mechanism, Polya urn is widely used in Bayesian analysis (see for example \cite{Johnson} and \cite{Muliere}): the initial composition of the urn can be considered as the prior knowledge about the occurrence of an event; the sampling of the balls represents the outcome of the statistical experiment; and the reinforcement of balls can be seen as the Bayesian updating, whose strength is given by $s$, of the prior knowledge.\\
Polya urns have a lot of interesting properties. For example, thanks to exchangeability and de Finetti's representation theorem, conditionally on a random variable $\Theta$, the variables $X_t$ are i.i.d $Bernoulli(\Theta)$. Moreover, it can be shown that  $\Theta\sim Beta\left(w_0/s,b_0/s \right)$ and $Z_t\to_{a.s.}\Theta$. These are the essential properties for our approach to extreme shock models. We refer to \cite{Johnson} and \cite{Mahmoud} for a much more complete treatment.\\
Recently, in \cite{Muliere}, a generalization of the Polya urn scheme has been introduced under the name of reinforced urn process (RUP). Here we directly introduce our special case, referring to the next section for a more general setting.\\
Imagine to have a discrete set of states $v_0<v_1<v_2<...$ and associate to each of them a Polya urn $U(v_i)$, $i\geq0$. Every urn has its own initial composition $C(v_i)=(w(v_i), b(v_i))$, for $i\geq0$.\\
A reinforced urn process $\{X_n \}$ on $\{v_0,v_1,... \}$ is defined iteratively as follows: set $X_0=v_0$ and, for $n\geq1$, if $X_{n-1}=v_i$ sample urn $U(v_i)$. If the sampled ball is white, then Polya reinforce urn $U(v_{i})$ with $s>0$ white balls and set $X_n=v_{i+1}$. Otherwise, whether the ball is black, add $s$ black balls and set $X_t=v_0$. In other words, starting from urn $U(v_0)$ we sequentially sample from urns in $v_0,v_1,v_2,...$ until a black ball is chosen, then we start again the sequential sampling.\\
Now, let black balls represent the event ``extreme shock", while white balls indicate the event ``no shock or weak shock". Moreover, imagine for simplicity that the states $v_0<v_1<v_2<...$ are just time instants, such that $v_i=i$ for $i\geq0$ (please notice that this simplification is not necessary for the model to work and the states $v_i$ can be of every kind).\\
Repeating the iterative sampling several times, we obtain a sequence like this as realization:
\begin{equation}
\begin{split}
\{X_n\}_{n\geq1}&=\{v_0,v_1,v_2,v_0,v_1,v_0,v_1,v_2,v_0,v_1,...,v_{10},v_{11},v_0,.... \} \\
&=\{0,1,2,0,1,0,1,2,0,1,...,10,11,0,.... \}
\end{split}
\end{equation}
In other words, every time a black ball is sampled, a so called 0-block is created, i.e. a sequence of states between two different 0's.\\
Now, assume we have several different systems, say $k\geq1$, that are subjects to shocks as in the standard extreme shock model setting. Let us hypothesize that these systems are exchangeable, i.e. it is not relevant, for our analysis, the order in which we consider them. If the life of every system is represented by one of the 0-blocks, we have that the process $\{X_n\}$ is partially exchangeable in the sense of Diaconis and Freedman \cite{Diaconis}. In other words $\{X_n\}$ can be expressed as a mixture of Markov chains. In fact, it suffices to notice that by assumption the 0-blocks are exchangeable (every block represents the life of a system) and every block is a Markov chain, since, within each block, the probability of moving from one state to the other only depends on the last visited state.\\
Let $\xi_i$ be the last coordinate of the $i$-th 0-block, for $i=1,...,k$, of the process $\{X_n\}$. Since every $\xi_i$ is a function of the corresponding 0-block $i$, and the blocks are exchangeable, then the sequence $\{\xi_i\}$ is exchangeable as well.\\
Now, in order to guarantee that the process $\{X_n\}$ is recurrent, i.e. $P[X_n=0 \text{ for infinitely many }n]=1$, let us assume that
\begin{equation}
\lim_{n\to \infty}\prod_{i=1}^n\frac{b(v_i)}{w(v_i)+b(v_i)}=0.
\end{equation}
In this way we are sure that infinitely many 0-blocks can be created and we can thus define the sequence $\{\tau_n \}$ of stopping times, with $\tau_0=0$ and $\tau_n=\inf\{n>\tau_{n-1}:X_n=0\}$. Naturally, we have that $\xi_n=X_{\tau_{n-1}}$.\\
Let us finally enter in the Bayesian interpretation of extreme shock models using reinforced urn processes. Consider the first system subject to shocks. All the information we have about the probability of an extreme shock in state $v_0=0$ is just represented by the a priori information about the initial composition of urn $U(v_0)$ (remember that an extreme shock is represented by the extraction of a black ball). Hence we have that
\begin{equation}
P(\xi_1=0)=\frac{b(0)}{w(0)+b(0)} \qquad \text{and} \qquad
P(\xi_1>0)=\frac{w(0)}{w(0)+b(0)}.
\end{equation}
If a black ball is sampled and the system fails, then urn $U(0)$ is reinforced with $s$ extra black balls. Otherwise we add $s$ extra white balls. For clarity we call the new reinforced urn $U^1(0)$. From now on $U^i(v)$ and $C^i(v)$ represent the Polya urn and its composition in state $v$ after the $i$-th sampling.\\
If the system does not fail in $v_0$, since we have extracted a white ball, we move to urn $U(v_1)$, and so on until we pick a black ball. View that we are considering the first system and we do not have any information about past experiments, we simply obtain
\begin{equation}\label{proc}
P(\xi_1=v_r)=\frac{b(v_r)}{w(v_r)+b(v_r)}\prod_{j=0}^{r-1}\frac{w(v_j)}{w(v_j)+b(v_j)}.
\end{equation}
Now imagine that the first system fails in $v_r$, i.e. at time instant $r$, then we have that $C^1(v_r)=(w(v_r),b(v_r)+s)$, $C^1(v_j)=(w(v_j)+s,b(v_j))$ for $j=0,...,r-1$ and $C^1(v_j)=C^0(v_j)$ for $j>r$.\\
At this point we start considering the second system. We begin from the urn in state $v_0$, that now has been updated according to the history of the first system, being $U^1(v_0)$. Up to state $v_r$ we will use the information we have acquired about system one. From $v_{r+1}$ on we will come back to the initial prior information, since no history has been observed. These iterative procedure is repeated for all the systems considered. We can then state the following result.
\begin{proposition}
Given a reinforced urn process $\{X_n\}$ as described above, after observing $m$ systems, we have that the probability that the $(m+1)$-th system fails in $v_r$ is equal to
\begin{equation}\label{pred1}
P(\xi_{m+1}=v_r|\xi_1,...,\xi_m)=\frac{b(v_r)+sd_r}{w(v_r)+sf_r+b(v_r)+sd_r}\prod_{j=0}^{r-1}\frac{w(v_j)+sf_j}{w(v_j)+sf_j+b(v_j)+sd_j},
\end{equation}
while
\begin{equation}\label{pred2}
P(\xi_{m+1}>v_r|\xi_1,...,\xi_m)=\prod_{j=0}^{r}\frac{w(v_j)+sf_j}{w(v_j)+sf_j+b(v_j)+sd_j},
\end{equation}
where 
$f_l=\sum_{p=1}^m1_{\{v_l<\xi_p \}}$ and $d_l=\sum_{p=1}^m1_{\{v_l=\xi_p\}}$.
\end{proposition}
\begin{proof}
The proof is simply given by applying the iterative construction of the process as for equation (\ref{proc}) and by counting the number of defaults in the different states at every sampling.
\end{proof}\\ \\
As far as inference is concerned, from equations (\ref{pred1}) and (\ref{pred2}), we derive that 
\begin{equation}
\hat{\xi}_{m+1}=E[\xi_{m+1}|\xi_1,...,\xi_m]=\sum_r v_rP(\xi_{m+1}=v_r|\xi_1=a_1,...,\xi_m=a_m).
\end{equation}
Anyway, the most interesting aspect of the reinforced urn process construction for extreme shock models is represented by the following proposition, that gives useful information about the distribution of default times.
\begin{proposition}
If $\{X_n \}$ is a recurrent reinforced urn process, the sequence $\{\xi_n\}$ is exchangeable, i.e. there exists a random distribution function $F$ such that, conditionally on $F$, the random variables of the sequence $\{\xi_n\}$ are i.i.d. with distribution $F$, whose law is that of a beta-Stacy process. In other words, for $v_{i+1}\in V$, the random mass assigned by $F$ to the subset $\{v_0,v_1,...,v_i+1\}$ is equal to
\begin{equation}
1-\prod_{j=1}^{i+1}(1-Y_j),
\end{equation}
where $Y_i$, $i=1,2,...$, are independent random variables such that $Y_i$ is distributed as $Beta(w(v_i)/s,b(v_i)/s)$.
\end{proposition}
\begin{proof}
The proposition is an application of Theorem 3.26 in \cite{Muliere}, to which we refer for further details. The fact that $Y_i$, $i=1,2,...$, are independent $Beta(w(v_i)/s,b(v_i)/s)$ random variables is a direct consequence of the use of Polya urns.
\end{proof}\\ \\
The beta-Stacy process has been introduced in \cite{Walker} as a special case of neutral to the right processes (see \cite{Doksum}). An interesting property of beta-Stacy processes is conjugacy, that is to say that, in our construction, the posterior distribution for $F$, after observing the history of the different systems, is still a discrete time beta-Stacy process, with jumps in $\{v_r\}$ and parameters $(w(v_r)+sf_r,b(v_r)+sd_r)$.\\\\
\textbf{Example}\\
We now show a possible simple application of the urn-based shock model in the field of material science and engineering.\\
Imagine we have several equal bars of the same metal and we want to test which is the maximum load that they can bear before rupture. A similar situation can be efficiently modeled using the reinforced urn process we have just described.\\
Set $v_0<v_1<v_2<...$ to be different loadings in increasing order. Then equip every state $v_i$ with an urn $U(v_i)$ as seen before.\\
Start with the first bar and apply loading $v_0$. If it cracks then read this as the extraction of a black ball from urn $U(v_0)$, reinforce the urn with $s$ black balls and pass to the second bar. If the loading is not sufficient for the bar to break, we have sampled a white ball, we reinforce the urn and then move to loading $v_1>v_0$. Finally we go on until the metal bar shatters.\\
Repeating the procedure for the different bars we are able to make Bayesian inference about the loading capacity of the different bars (remember equations (\ref{pred1}) and (\ref{pred2})). Calibrating the magnitude $s$ of reinforcement, we can weight the information given to the experiment with respect to our prior knowledge, as expressed by the initial compositions of the different urns.\\
This kind of model can be considered a special case of extreme shock model, in which the magnitude of shocks (loading) is not random, but it increases state after state.

\section{The general construction and generalized extreme shock models}

The model we have introduced for extreme shook models is just one of the possible application of RUP to shock models. For example, if we move to a more general definition of RUP, in which every urn needs not to be a Polya urn, we can obtain several interesting results. For example, we can reproduce the urn-based generalized extreme shock model of \cite{Cihu}.\\
A generalized reinforced urn process, i.e. a generalization of the construction in \cite{Muliere}, can be obtained using the following ingredients:
\begin{enumerate}
\item A countable state space $V$. 
\item A finite set $E$ of colors, whose cardinality is at least 1.
\item For every $v\in V$ there exists an urn $U(v)$ characterized by an initial composition $C(v)$ and a reinforcement matrix $M(v)$. In particular $C(v)=\{n_{v}(c):c\in E \}$ where $n_{v}(c)\geq0$ represents the number of balls of color $c$ in $U(v)$; and $M(v)$ describes how the urn behaves when it is sampled, i.e. how balls are replaced or reinforced.
\item A function $m:V \times E \rightarrow V$ that indicates how the process moves from state to state, that is from one urn to the other.
\end{enumerate}
Imagine we are in $v\in V$. We sample the urn $U(v)$ and look at the color $c$ of the extracted ball, whose probability of being sampled depends on the composition $C(v)$ of the urn. At this point the matrix $M(v)$ says how we have to behave (see \cite{Cihu} and \cite{Mahmoud} for more details on reinforcement matrices). For example if $M(v)$ is the null matrix, it could represent sampling without replacement, so that the sampled ball is thrown away. If $M(v)$ is the identity matrix, then we consider sampling with replacement. A Polya urn is represented by a diagonal matrix having $1+s$, with $s>0$, on the main diagonal. And so on for all the other possible schemes. At this point the law of motion $m$ says how we move form state $v_1\in V$ to say state $v_2\in V$, i.e. $m(v_1,c)=v_2$.\\
To clarify the exposition, let us consider the simple model we have introduced before. Set $\{X_{t}\}$ to be our process defined on $V$. Assume $V=\{0,1,2,...\}$, $E=\{b, w\}$ where $b$ means black and $w$ stands for white, $C(v)=[w(v),b(v)]$ and $M(v)=\left[\begin{array}{cc}1+s & 0 \\0 & 1+s\end{array}\right]$, $s>0$, for every $v\in V$. Moreover the law of motion $m$ is such that $m(v,b)=0$ and $m(v,w)=v+1$.\\
For what concerns generalized extreme shock models, as we have seen in \cite{Cihu} the authors propose a generalized triangular Polya-like urn, in order to model the risky threshold mechanism introduced in \cite{Huesler}.\\
It is not difficult to reproduce the results in \cite{Cihu} using the generalized reinforced urn process. In fact it suffices to mimic the behavior of the original triangular urn using our sequence of urns. The trick is to introduce a recurrence relation among the urns in order to imitate the evolution of the urn-based generalized extreme shock model.\\
Set $V=\{0,1,2,...\}$ and $E=\{w,r,b\}$. For every $v\in V$ define the balanced matrix
\begin{equation}
M=M(v)=\left[
\begin{array}
[c]{ccc}%
1+s & 0 & 0\\
0 & 1+r & p\\
0 & 0 & 1+s
\end{array}
\right].  
\end{equation}
Assume that the vector representing the initial composition of urn $U(0)$ is $C(0)=(a_0,b_0, c_0)$ and, for every $v>0$, set
\begin{equation}
C(v)=C(v-1)+e_{v-1}M,
\end{equation}
where $e_{v}$ is a random vector equal to $\left[\begin{array}{ccc}1 & 0 & 0\end{array}\right]$ if the ball extracted in $U(v)$ is white, $\left[\begin{array}{ccc}0 & 1 & 0\end{array}\right]$ if red and $\left[\begin{array}{ccc}0 & 0 & 1\end{array}\right]$ if black. Furthermore, for the law of motion, simply set $m(v,w)=m(v,r)=v+1$ and $m(v,b)=0$.\\
It is effortless to verify that the so-defined urn chain exactly reproduces the urn model introduced in \cite{Cihu}, even if in a less intuitive way. As a consequence of this, all the results there stated still hold.

\section{Conclusions}
We have seen that extreme shock models are models in which a system is subject to random shocks of random magnitude, which make it fail as soon as they overcome a certain resistance threshold. Extreme shock models can have several applications in material science (as in the simple example we have shown) and engineering in general, but also in economics and finance, think for example about firms' defaults. \\
In this paper we have proposed an alternative modeling based on a special version of the reinforced urn processes introduced by \cite{Muliere}. The main novelty of this approach is given by the possibility of performing a Bayesian nonparametric analysis of extreme shocks. Given a set of systems subjects to shocks, and assuming that they are exchangeable, we have shown how to exploit the information about their history for making predictions concerning default times. Moreover, the use of Polya urns for the construction of the model allows the researcher to introduce his/her prior knowledge of the phenomena, by modifying the initial composition of the different urns.\\
Possible extensions of the present work could be its development in continuous time, on the basis of the main results in \cite{Muliere2}, and its applications to practical problems, such as for example fatigue analysis \cite{Stephens}. Moreover, it could be also worth of investigation to work with the general definition of reinforced urn processes, introducing for example time-variant reinforcement matrices, in order to model other shock models as discussed in \cite{Huesler}.
\\
\\
\textbf{Acknowledgements:} The present work has been supported by the Swiss National Science Foundation.

\end{document}